\documentclass[twoside,11pt]{report}
\usepackage{graphicx}
\usepackage{amssymb, amsmath, amsxtra}
\usepackage{epsfig,subfigure}

\usepackage{latexsym}
\usepackage{bm}
\usepackage{wrapfig}

 \usepackage{ntheorem}
           \theoremstyle{plain}
                      {\theorembodyfont{\rmfamily}
                      \theoremseparator{.}
                       
           \newtheorem{theorem}{Theorem}[section]
           \theoremstyle{plain}
           
           \theoremstyle{plain} 
           \theoremstyle{plain}

           \theoremstyle{plain}
           \newtheorem{remark}{Remark}[section]}

\begin{document}

\today

\begin{center}
{\Large \bf Pricing options in illiquid markets:\\ optimal
systems, symmetry reductions and exact solutions}\\[2ex]
 {\large  L. A. Bordag}\\[1ex]
{\it IDE, MPE Lab,\\ Halmstad University, Box 823,\\ 301 18
Halmstad, Sweden}
 \end{center}

 \noindent
{\bf Abstract}\\
We study a class of nonlinear pricing models which involves the
feedback effect from the dynamic hedging strategies on the price
of asset introduced by  Sircar and Papanicolaou. We are first to
study the case of a nonlinear demand function involved in the
model. Using a Lie group analysis we investigate the symmetry
properties of these nonlinear diffusion equations. We provide the
optimal systems of subalgebras and the complete set of
non-equivalent reductions of studied PDEs to ODEs. In most cases
we obtain families of exact solutions or derive particular
solutions to the equations.\\[2ex]
[5pt]

\section{Introduction}
\label{intr}
 \setcounter{equation}{0}
 One of the important assumptions of the
 classical Black-Scholes theory is the assumptions that any
 trading strategy of any trader on the market do not affect asset
 prices. This assumption is failed in the presence of large
 traders whose orders involve a significant part of the available
 shares. Their trading strategy has a strong feedback effect on
 the price of the asset, and from there back onto the price of
 derivative products. The continuously increasing volumes of financial markets
 as well as a significant amount of large traders acting on
 these markets force us to develop and to
 study new option pricing models.\\
There are  a number of suggestions on how to incorporate in  a
mathematical model the feedback effects which correspond to
different types of frictions on the market like illiquidity  or
transaction costs. Most financial market models are characterized
by  nonlinear partial differential equations (PDEs) of the
parabolic type. They contain usually a small  perturbation
parameter $\rho$ which vanishes if the feedback effect is removed.
If $\rho$ tends to zero then the corresponding nonlinear PDE tends
to the Black-Scholes equation.

Some of the option pricing models in illiquid markets possess
complicated analytical and algebraic structures which are singular
perturbed. We deal with singular perturbed PDEs if one of the
nonlinear terms in the studied equation incorporates the highest
derivative multiplied by the small parameter $\rho$. It is a
demanding task to study such models. Solutions to a singular
perturbed equation may blow up in the case $\rho=0$ and may not
have any pendants in the linear
case.\\
 An example of a singular perturbed model is the continuous-time model developed by Frey
 \cite{bib:frey-98a}. He derived a PDE for perfect replication
 trading strategies and option pricing for the large traders. An
 option price $u(S,t)$ in this case is a solution to the nonlinear
 PDE
 \begin{equation} \label{frey}
 u_t + \frac{1}{2} \frac{\sigma^2  S^2 u_{SS}}{\left(1 - \rho
  \lambda(S) S u_{SS}\right)^2}  =0 \,,
\end{equation}
where $t$ is time, $S$ denotes the price and $\sigma$ the
volatility of the underlying asset. The continuous function
$\lambda(S)$ included in the adjusted diffusion coefficient
depends on the payoff of the derivative product. The Lie group
analysis and properties of the invariant solutions to Eq.
(\ref{frey}) for different types of the function $\lambda(S)$ were
studied in \cite{Bordag}-\cite{BordagFrey}. The analytic form of
the invariant solutions to this model allow us to follow up the
behavior of these solutions.

Under the similar assumptions  Cetin, Jarrow and Protter
\cite{Protter} developed a model which includes liquidity risk for
a large trader. Liquidity risk is the additional risk due to the
timing and size of trade. The value $u(S,t)$ of a  self financing
trading strategy for the large trader in this setting is a
solution of the following nonlinear PDE
\begin{equation}
u_t+\frac{1}{2}\sigma^2 S^2 u_{SS}(1-\rho S u_{SS})^2=0.
\label{jarrowProtter}
\end{equation}
This equation seems to be simpler as the previous one but it is
still  a singular perturbed PDE. The Lie group analysis and the
symmetry algebra admitted by this equation were studied in
\cite{Bobrov}.\\

Sircar and Papanicolaou in \cite{bib:papanicolaou-sircar-98}
present a class of nonlinear pricing models that account for the
feedback effect from the dynamic hedging strategies on the price
of asset using the idea of a demand function of the reference
traders relative to the supply. They obtain a nonlinear PDE of the
following type
\begin{equation}\label{sircpapgen}
u_{t}+\frac{1}{2}\left[\frac{U^{-1}(1-\rho u_{S})U'(U^{-1}(1-\rho
u_{S}))}{U^{-1}(1-\rho u_{S})U'(U^{-1}(1-\rho u_{S}))- \rho S
u_{SS}}\right]^2\sigma^2S^2 u_{SS}+r(S u_{S}-u)=0,
\end{equation}
where $t$ is time, $S$ and $\sigma$ is the price and the
volatility of the underlying asset respectively, and the parameter
$r$ is the risk-free interest rate. The value $u(S,t)$ is the
price of the derivative security and depends on the form of the
demand function $U(\cdot)$. The expression $U^{-1}(\cdot)$ denotes
the correspondingly inverse function, because of the strong
monotonicity of the demand function the existence of the inverse
function $U^{-1}(\cdot)$ is guaranteed. In the bulk of their paper
\cite{bib:papanicolaou-sircar-98} authors studied the particular
model arising from taking $U(\cdot)$ as linear, i.e. $U(z)=\beta
z, ~ \beta>0 $. The authors mainly focused on the numerical
solution and discuss the difference to the classical Black-Scholes
option pricing theory.

In the present paper we study a more general case in which the
demand function of the type $U(z)=\beta z^{\alpha}, ~\alpha,\beta
\ne 0 $ is incorporated. Consistency of (\ref{sircpapgen}) with
the Black-Scholes model characterizes the class of the admitted
demand functions and leads to the condition $U'(z)=\beta \alpha
z^{\alpha-1}>0$.
 In this case the model (\ref{sircpapgen}) takes the form
\begin{equation} \label{sipar}
\frac{\partial u}{\partial t}+\frac{1}{2}\left[\frac{1-\rho
\frac{\partial u}{\partial S}}{1-\rho \frac{\partial u}{\partial
S}-\frac{\rho}{\alpha} S\frac {\partial^2 u}{\partial
S^2}}\right]^2\sigma^2S^2\frac{\partial^2 u}{\partial
S^2}+r\left(S\frac{\partial u}{\partial S}-u\right)=0.
\end{equation}
The diffusion coefficient in Eq. (\ref{sipar}) depends on both,
$u_S$ and $u_{SS}$ multiplied by the small perturbation parameter
$\rho$. It depends also on  the parameter $ \alpha $ which
characterizes the type of the demand function and on the interest
rate $r$.

 We give a short overview of analytical properties of
Eq. (\ref{sipar}) in the next Section 2. In Section 3 we provide
the Lie group analysis of this equation. Depending on whether the
interest rate  $r=0$ or $r \ne 0$, we obtain different Lie
algebras admitted by the respectively equation. Then we provide
optimal systems of subalgebras in the both cases. The optimal
systems of subalgebras give us the possibility to describe the set
of independent reductions of these nonlinear PDEs to different
ordinary differential equations (ODEs). In some cases we found the
explicit solutions to these equations. We discuss the properties
of invariant solutions in Section 4.

\section{Basic analytical properties of Eq. (\ref{sipar})}
 \label{analyt}
 \setcounter{equation}{0}
In the model  (\ref{sipar}) introduced by Sircar and Papanicolaou
in \cite{bib:papanicolaou-sircar-98} the diffusion coefficient has
a very complicated analytical and non trivial algebraic structure.
In particular the diffusion term is represented by a fraction
which contains derivatives $u_S(S,t), u_{SS}(S,t)$. The authors
analyzed some analytical properties of this equation in the case
$\alpha =1$ in the vicinity of the Black-Scholes equation and
considered a valuation of a European derivative security  with a
convex payoff using their model. They give asymptotic results for
when the volume of assets traded by the large traders is small
compared to the total number of unites of
the asset.\\
In this Section we discuss some global analytical properties of
Eq. (\ref{sipar}) for $\alpha \ne 1$.

We pay the main attention to the second term in (\ref{sipar}).
 We assume that the space variable $S \in \Omega \cup \{
0\}$, where $\Omega = \mathbb{ R}^+ $ and the time variable $t$
lie in $ {\cal T}\cup \{ 0\}$, where ${\cal T}= \mathbb{R}^+ $.
This term can vanish for some values of the variable $S$ or on
some set of smooth functions and then the equation may change the
type from the parabolic one to another one. Other hand the
fraction in the second term may became meaningless because of
vanishing of the denominator on some set of smooth functions. We
should exclude such functions from the domain
of definition of our model.\\
The classical linear diffusion equation of type $u_t= u_{SS}$ is
well defined on the space $D= C^{2,1}(\Omega \times {\cal T})
\bigcap C(\{\Omega \cup  \{0\} \} \times \{ {\cal T}\cup \{0\}
\})$ and $u(S,t)$ map the space $D$ to a space of continuous
functions $M=C(\{\Omega \cup  \{0\} \} \times \{ {\cal T}\cup
\{0\} \}).$

Let us check whether the expression for the diffusion coefficient
in Eq. (\ref{sipar})  vanishes or has
singularities.\\
 Fist we study the case that the denominator of the
 fraction in (\ref{sipar}) is equal to zero, i.e. we have to
 solve the equation
\begin{equation} \label{sipaDenominator}
1-\rho u_S - \frac{ \rho}{\alpha} S  u_{SS}  =0.
\end{equation}
 It is easy to see that this equation has the following solution
\begin{eqnarray}
u_{sing}(S,t)= \frac{S}{\rho} + c_1(t)\frac{\alpha}{\alpha-1}
S^{\frac{\alpha-1}{\alpha}} +c_2(t),~~~ \alpha \ne 1, \nonumber \\
u_{sing}(S,t)= \frac{S}{\rho} + c_1(t)\ln (S) +c_2(t),~~ \alpha
=1,\label{solsing}
\end{eqnarray}
where $c_1(t)$ and $c_2(t)$ are arbitrary functions of $t$. We can
rewrite the expressions in (\ref{solsing}) as one expression which
includes the case $\alpha =1$ as a limit case. Then we obtain
\begin{equation}
u_{sing}(S,t)= \frac{S}{\rho} + c_1(t)\frac{\alpha}{\alpha-1}
\left(S^{\frac{\alpha-1}{\alpha}} -1 \right) +c_2(t).
\label{singtogether}
\end{equation}
The numerator of the second term in (\ref{sipar}) is equal to zero
if one of the equations is satisfied
\begin{eqnarray}
S^2 u_{SS}=0, \nonumber \\
1-\rho u_S =0. \label{numer}
\end{eqnarray}
The first equation is satisfied on all linear functions of $S$ and
in the point $S=0$, the second equation has the following solution
\begin{equation}\label{numer0}
u_{0}(S,t)= \frac{S}{\rho} + c_2(t).
\end{equation}
We notice that in the case $c_1(t)=0$ the functions
$u_{sing}(S,t)$ and $u_{0}(S,t)$ coincide. It means in this case
the numerator and the denominator of the fraction in the equation
(\ref{sipar}) are simultaneously equal to zero.\\

In the second step we should define a limiting procedure to
explain what we means if we say that (\ref{numer0}) is a solution
to (\ref{sipar}).

We chose in the space $D$ a one-parametric family of functions
$u_{\epsilon}(S,t)$ of the following type
\begin{equation}
u_{\epsilon}(S,t)=d_1(t) S + d_2(t) +\epsilon v(S,t),
\label{oneparam}
\end{equation}
where $ \epsilon \in \mathbb{R}$ is a parameter, the functions
$d_1(t), d_2(t)$ are arbitrary functions of time and $v(S,t) \in
D$. If now the parameter $\epsilon \to 0$ then the family of
functions of the type (\ref{oneparam}) converges in the norm of
the space $D$ to a linear function of $S$, i.e. to
$u_{0}(S,t)=d_1(t) S + d_2(t)$. We apply to this family the
differential operator defined by (\ref{sipar})
\begin{eqnarray} \label{ner}
d_1^{'}(t) S + d_2^{'}(t) + \epsilon v_t (S,t)+ \frac{\sigma^2}{2}
\frac{\epsilon S^2 v_{SS}(1- \rho d_1(t)- \epsilon \rho v_S)^2
}{\left(1- \rho d_1(t)- \epsilon \rho v_S - \epsilon \beta \rho
   S v_{SS}\right)^2} =0 \,,
\end{eqnarray}
here $d_1^{'}(t), d_2^{'}(t) $ denotes the first derivatives of
the corresponding functions. From (\ref{ner}) follows that any
linear function of $S$ with constant coefficients will be solution
to equation (\ref{sipar}) if the last term in (\ref{ner}) is a
bounded function in the norm of the space $M$. If we replace the
linear part in (\ref{ner}) by $u_0(S,t)$ we see that this function
(\ref{numer0}) is a solution to (\ref{sipar}) if $c_2(t)= const.$.
The fraction in (\ref{ner}) is not well defined just in one case
if $v(S,t)$ coincide with the second term in (\ref{singtogether}).
So far we use as the domain of definition for our model the space
$D$ the functions of type (\ref{singtogether}) with $c_1(t) \ne 0$
are excluded because they or their derivatives have singularities
in the point $S=0$ and consequently they do not belong to the
space $D$.

In the classical case of a linear parabolic diffusion equation
solutions of type (\ref{singtogether}) which do not belongs to the
set of classical solutions are called viscosity solutions
\cite{Barles}, \cite{Grandal} and these solutions are well
studied.

We proved that the functions of type (\ref{singtogether}) should
be excluded from the further investigation because the model
(\ref{sipar}) is not well defined on them. The linear function
(\ref{numer0}) with $c_2(t)=const.$ is a solution to (\ref{sipar})
because any one parametric family of functions $u_{\epsilon}(S,t)$
in the norm of the space $D$ convergent to $u_{0}$ is mapped by
the differential operator defined by (\ref{sipar}) to a
zero-convergent family of functions in the norm of the space $M$.

\section{Symmetry properties of the model}
 \label{sym}
 \setcounter{equation}{0}
We provide in this Section the Lie group analysis of Eq.
(\ref{sipar}) first for the case $r=0$ then for $r \ne 0$. In both
cases it is possible to find  the non-trivial Lie algebras
admitted by the equation. We use the standard method to obtain the
symmetry group suggested by Sophus Lie and developed further in
\cite{{Ovsiannikov}}, \cite{{Olver}} and \cite{Ibragimov}. In the
case $r=0$ we obtain a four dimensional Lie algebra $L_4$ and by
$r \ne 0 $ Eq. (\ref{sipar}) admits a three dimensional algebra
$L_3$ defined in the subsection 3.2.

 All three and
four dimensional real Lie algebras and their subalgebras were
classified by Pattera and Winternitzs in
\cite{PateraWinternitzs1977}. The authors looked for
classifications of the subalgebras into equivalence classes under
their group of inner automorphisms. They used also the idea of
normalization which guarantees that the constructed optimal system
of subalgebras is unique up to the isomorphisms.

The symmetry group $G_4$ related to the symmetry algebra $L_4$ is
generated by a usual exponential map. We use the similar procedure
to obtain to each subalgebra $h_i$ from the optimal system of
subalgebras the correspondingly subgroup $H_i$.

The optimal system of subalgebras allows us to divide the
invariant solutions into non-intersecting equivalence classes. In
this way it is possible to find the complete set of essential
different invariant solutions to the equation under consideration.

Using the invariants of these subgroups  we reduce the studied PDE
to different ODEs. Solutions to these ODEs give us the invariant
solutions to the nonlinear PDE (\ref{sipar}) in an analytical
form. In the both cases whether by $r=0$ or by $r\ne0$ we skip the
study of invariant reductions to the two- and three-  dimensional
subgroups because of they give trivial results for the studied
equation.

 \subsection{Symmetry reductions in the case $r=0$}

In the first step we solve the Lie determining equations for the
equation
\begin{equation} \label{sipa}
u_t + \frac{1}{2} \frac{\sigma^2(1-\rho u_S)^2 }{\left(1-\rho u_S
- \frac{\rho}{\alpha}
   S u_{SS}\right)^2} S^2 u_{SS}  =0 \,, {\alpha} \ne 0,
\end{equation}
and obtain the Lie algebra admitted by this equations. We
formulate the results in the following theorem.

{\begin{theorem} Eq. (\ref{sipa}) admits a four dimensional Lie
algebra $L_4$ with the following infinitesimal generators
\begin{equation}
e_1=- \frac{S}{2}\frac{\partial}{\partial S} +\left(\frac{S}{2
\rho} -  u \right)\frac{\partial}{\partial u},
~~e_2=\frac{\partial}{\partial u}, ~~e_3=\frac{\partial}{\partial
t},~~~ e_4=\rho S \frac{\partial}{\partial S}+ S
\frac{\partial}{\partial u}. \label{generatorsSiPa}
\end{equation}
The commutator relations are
\begin{eqnarray}
[e_1,e_3]=[e_1,e_4]=[e_2,e_3]=[e_2,e_4]=[e_3,e_4]=0,~
[e_1,e_2]=e_2,
\end{eqnarray}
\end{theorem}
\begin{remark} In the very short letter \cite{Bordag:2008} there is a misprint in the theorem
formulation. We apologize by readers for the inconvenience.
\end{remark}
The Lie algebra $L_4$ has a two-dimensional subalgebra
$L_2=<e_1,e_2>$ spanned by the generators $e_1,e_2$. The algebra
$L_4$ is a decomposable Lie algebra and can be represented as a
semi-direct sum
 $L_4=L_2 \bigoplus e_3 \bigoplus e_4$.
The optimal system of subalgebras for $L_4$ were provided in
{\cite{PateraWinternitzs1977} and presented in Table \ref{optsev}.

\begin{table}
 {\begin{tabular}{|c|c|}
\hline
{Dimension}&{Subalgebras}\\
\hline
$1$&$h_1=<e_2>,~~h_2=<e_3 \cos {(\phi)}+e_4\sin{(\phi)}>,$\\
&$h_3=<e_1+ x(e_3 \cos {(\phi)}+e_4\sin{(\phi)})>,$\\
&$h_4=<e_2 +\epsilon(e_3 \cos{(\phi)}+e_4\sin{(\phi)})>$\\
\hline
$2$&$h_5=<e_1+x(e_3 \cos {(\phi)}+e_4\sin{(\phi)}),e_2>,~~h_6=<e_3,e_4>,$\\
&$h_7=<e_1 +x(e_3 \cos {(\phi)}+e_4\sin{(\phi)}),e_3 \sin{(\phi)}-e_4\cos{(\phi)}>,$\\
&$h_8=<e_2 +\epsilon(e_3 \cos{(\phi)}+e_4\sin{(\phi)}),e_3 \sin{(\phi)}-e_4\cos{(\phi)}>,$\\
&$h_9=<e_2,e_3 \sin {(\phi)}-e_4\cos{(\phi)}>$\\
\hline
$3$&$h_{10}=<e_1,e_3,e_4>,~~h_{11}=<e_2,e_3,e_4>,$\\
&$h_{12}=<e_1 +x(e_3 \cos {(\phi)}+e_4\sin{(\phi)}),e_3 \sin {(\phi)}-e_4\cos{(\phi)},e_2>$\\
\hline
\end{tabular}}
\caption{\cite{PateraWinternitzs1977} The optimal system of
subalgebras $h_i$ of the algebra $L_4$ where $x\in {\mathbb
R},~~\epsilon=\pm1,~~\phi \in [0,\pi ] $.} \label{optsev}
\end{table}

The optimal system of the one-dimensional subalgebras involves
four subalgebras $h^0_i,~i=1,\dots,4$. We take step-by-step each
of these subalgebras $h^0_i$ and the corresponding symmetry
subgroup $H^0_i$ and study which invariant reductions of the studied  PDE are possible.\\

{\bf Case $H^0_1$.} This one-dimensional subgroup $H^0_1 \subset
G_4$ is generated by the subalgebra $
h^0_1=<e_2>=<\frac{\partial}{\partial u}> $. It means that we deal
with a subgroup of translations in the $u$ - direction. Hence, to
each solution to Eq. (\ref{sipa}) we can add an arbitrary constant
without destroying the property of the function to be solution.
This subgroup does not provide any
reduction.\\

{\bf Case $H^0_2$.} The subalgebra $h^0_2$ is spanned by the
generator $e_3 cos(\phi) +e_4 sin(\phi)$. In terms of the
variables $S,t,u$ it takes  the form
\begin{equation} \label{h2inr0}
h^0_2= <\frac{\partial}{\partial t}\cos (\phi)+\left( \rho S
\frac{\partial}{\partial S}+ S \frac{\partial}{\partial u}\right)
\sin (\phi) >.
 \end{equation}
The invariants $z,w$ of the corresponding subgroup $H^0_2$ are
equal to
\begin{eqnarray} \label{invh2r0}
z=S\exp(-t \rho \tan(\phi) ),~~ w = u-\frac{1}{\rho}S
\end{eqnarray}
and we take them as the new dependent and independent variables,
respectively. Then the PDE (\ref{sipa}) is reduced to the ordinary
differential equation of the following form
\begin{equation} \label{redeqh2r0}
 - \rho {\delta} zw_z + \frac{1}{2}\sigma^2w_{zz}z^2\left(\frac{\alpha w_z}{\alpha
 w_z+zw_{zz}}\right)^2=0,~
 \delta=\tan(\phi), \phi \in [0,\pi], \phi \ne \frac{\pi}{2}.
 \end{equation}
This second order ODE is reduced to the first order equation by
the substitution $w_z=v(z)$ which takes the form
\begin{equation}\label{redFirsteqh2r0}
z v \left( - \rho \delta +\frac{\sigma^2}{2} z (\ln{v})_z \left(
\frac{\alpha}{z (\ln{v})_z +\alpha} \right)^2 \right)=0.
\end{equation}
 Eq. (\ref{redFirsteqh2r0}) has two trivial solutions $z=0$ and
$v(z)=w_z=0$ which are not very interesting for applications. The
non trivial solutions we obtain if we set the last factor in Eq.
(\ref{redFirsteqh2r0}) equal to zero. We obtain the solution to
(\ref{redFirsteqh2r0}) in the form
\begin{equation}\label{h2solr01}
w(z)= c_1 z^p, ~~ p=1+\alpha -a \pm \sqrt{a(a-2 \alpha)},~
a=\frac{\sigma^2}{4 \rho \delta},
\end{equation}
where $c_1$ is an arbitrary constant. In terms of the variables
$S,t,u$ the solution (\ref{h2solr01}) is equivalent to the
following solution to Eq.(\ref{sipa})
\begin{equation}\label{h2solr0}
u(S,t)= c_1 S^p \exp(-p \rho \tan(\phi) ~t), ~~ p=1+\alpha - a \pm
\sqrt{a(a - 2 \alpha)},~ a=\frac{\sigma^2}{4 \rho \delta},
\end{equation}
where $ \delta=\tan(\phi), \phi \in [0,\pi], \phi \ne
\frac{\pi}{2} $.\\

{\bf Case $H^0_3$.} The  subalgebra $h^0_3$ is spanned by
$$h^0_3=<x
\cos (\phi)\frac{\partial}{\partial t}+\left(x\rho\sin (\phi)-
 \frac{1}{2}\right)S\frac{\partial}{\partial S}+
 \left(\left(\frac{1}{2\rho}+x\sin(\phi)\right) S-u\right)\frac{\partial}{\partial
 u}>.$$
The invariants $z,w$ of the corresponding  subgroup $H^0_3$ are
given by the  expressions
\begin{equation}\label{invh3r0}
z=Se^{-ct},~~ w=S^b u-\frac{1}{\rho}S^{1+b},
\end{equation}
where $b=(x\rho\sin (\phi)-\frac{1}{2})^{-1},$ $c=({b
x\cos(\phi)})^{-1}$, $x \ne 0$, $\phi \in [0,\pi], \phi \ne
\frac{\pi}{2}$ and $x\rho\sin (\phi)-\frac{1}{2}\ne 0$.\\
We use $z,w$ as the new invariant variables and reduce
Eq.(\ref{sipa}) to the ODE
\begin{equation}\label{redh3r0}
- c z w_z +\frac{\sigma^2 \alpha^2}{2} \frac{\left( z^2 w_{zz} -2
{b} z w_z +{b}\left( 1+{b}\right)w \right) \left( z w_z - {b}w
\right)^2 } {\left(\alpha \left( z w_z - {b}w \right)+ \left( z^2
w_{zz} -2 {b} z w_z +{b}\left( 1+ {b}\right)w \right)\right)^2}=0.
\end{equation}
Eq. (\ref{redh3r0}) admits a solution of the type
\begin{equation} \label{solal}
w(z)=c_1 z^q,
\end{equation}
where $q$ is a real root of the polynomial of the degree 5
\begin{eqnarray}\label{rootredh3r0}
- c q {\left(
 q (q-1) + \left(\alpha-2 {b}\right) q +{b}\left( 1+
{b}-\alpha \right) \right)^2} \nonumber\\+\frac{\sigma^2
\alpha^2}{2} {\left( q (q-1) -2 {b}q +{b}\left( 1+{b}\right)
\right) \left( q - {b} \right)^2 } =0.\nonumber
\end{eqnarray}
Eq. (\ref{redh3r0}) has a complicate structure and is hardly
possible to solve it in the general form. But by a special values
of involved parameters we can simplify the equation and obtain
some particular classes of solutions.\\
 We take the special case of Eq. (\ref{sipa}) with $\alpha=1$.\\
 Under the special choice of the
parameters $\phi=0,\pi$ and $b=-2$ in Eq. (\ref{invh3r0}) and by
using  the invariants $z,w$ in the form
$$z= S \exp
\left(\frac{1}{2 x} b t \right), x\ne 0, ~~w= S^{-2} u - (\rho
S)^{-1}, $$ we reduce Eq.(\ref{sipa}) to the ODE
\begin{equation}\label{redfi}
z w_z ((z(z^2 w)_{z})_z)^2 + \sigma^2 x~~(z^2 w )_{zz} ((z^2
w)_z)^2=0.
\end{equation}
The substitution (\ref{solal}) in this case leads to the second
order algebraic equation on the value of the parameter $q$
\begin{equation}\label{reh1}
q^2 + q \left(2 +  \sigma^2 x \cos (\phi)\right) + \sigma^2 x \cos
\phi=0, ~~q \ne -2, \phi=0,\pi,
\end{equation}
which has two roots $q_1=- \sigma^2 x \cos (\phi)$ and $q_2=-2.$
For the future study we take just the first value $q_1$. The value
$q_2=-2$ leads to the know solution $u_0(S,t)$ (\ref{numer0}).
Since by $q_2=-2$ both, the numerator and the denominator in the
fraction in (\ref{sipa}) vanish, and the solution (\ref{solal})
coincide then with $u_{0}(S,t)$ by $c_2(t)=const.$ which we discussed in the previous section.\\

We notice that the solution (\ref{solal}) differs from the
function $u_{sing}(S,t)$ (\ref{solsing}) which involves the
logarithmic term in the case $\alpha=1$.

The solution to (\ref{redfi}) or respectively to Eq.(\ref{sipa})
in the form (\ref{solal}) in terms of $S,t,u$ variables is equal
to
\begin{equation} \label{solfirst}
u(S,t)=  \frac{S}{\rho} + C_1 S^{2- \sigma^2 x  }
e^{-\frac{\sigma^2}{2} t} + C_2 ,~~ \alpha \ne -2,
\end{equation}
where $C_1,C_2$ are arbitrary constants, $x \in \mathbb R$ and the
first term is the only term which contains the dependency on the
parameter $\rho$. We skip the factor $\cos (\phi)$ by $x$ in this
expression  because $\cos (\phi)=\pm 1$ in the case $\phi = 0,\pi$
and $x\in \mathbb R$.

It is remarkable that the reduced equation (\ref{redfi}) does not
contain any more the parameter $\rho$. Hence all invariant
solutions of this class can be represented as a sum of two terms:
the first one is equal to $S/{\rho}$ and the second one depends on
$z$ only but not on the parameter $\rho$.

 If we left the values of parameters like in the previous case, but take
 the invariants in another form
  $z=\ln{S}+t b/4 $ and $w(z)=(u/S - 1/{\rho})S^{\gamma}$
 than we obtain the different form of the reduced ODE
\begin{eqnarray}\nonumber
w_{zz}^2 w_z +w_{zz}(w^2_{z}(4(1-\gamma)+\kappa)+w_{z} w (2
(1-\gamma)^2 +2 \kappa (1-\gamma))\\ \nonumber +w^2 \kappa
(1-\gamma)^2 )+ w_z^3 (4(1- \gamma )^2+\kappa(1-2 \gamma)) +w_z^2
w (1-\gamma)(4 (1-\gamma )^2 \\+ \kappa (2 - 5 \gamma))
 + w_z w^2 (1-\gamma)^2
((1-\gamma )^2+\kappa (1-4 \gamma)) -\kappa \gamma (1-\gamma)^3
w^3=0,\label{seconred}
\end{eqnarray}
where $\kappa =2 \sigma^2/b$. This second order equation can be
reduced in the case $w_{z}\ne 0 $ to a first order ODE. We
substitute $p(w)=w_z(z(w))$ and correspondingly $w_{zz}=p_w p$,
i.e., $w$ is the independent variable and $p$ is the dependent
variable in this case. Then we obtain the first order ODE
\begin{eqnarray} \nonumber
p_w^2 p^3+ p_w p(p^2(4(1-\gamma)+\kappa)+p w (2 (1-\gamma)^2 +2
\kappa (1-\gamma)) +w^2 \kappa (1-\gamma)^2 )\\\nonumber + p^3
(4(1- \gamma )^2+\kappa(1-2 \gamma)) +p^2 w (1-\gamma)(4 (1-\gamma
)^2 + \kappa
(2 - 5 \gamma))\\
 + p w^2 (1-\gamma)^2
((1-\gamma )^2+\kappa (1-4 \gamma)) -\kappa \gamma (1-\gamma)^3
w^3=0.\label{seconredfirstod}
\end{eqnarray}
This equation is quadratic in the first derivative $p_w$ and it is
equivalent to the  system of two first order equations. For some
values of the constants $\gamma$ and $\kappa$ it can be explicitly
solved. The simplest case we obtain if we chose $\gamma =1$ then
the solution coincide with (\ref{solfirst}). In other cases the
equation can be studied using qualitative methods.

 {\bf Case $H^0_4$.} We
consider subalgebra $h^0_4$ spanned by
$$h^0_4=<\epsilon \cos
(\phi)\frac{\partial}{\partial t}+\epsilon\rho S \sin (\phi)
\frac{\partial}{\partial S}+
 (1+\epsilon S \sin (\phi))\frac{\partial}{\partial u}>.$$
The invariants $z,w$ are given by the expressions
\begin{eqnarray}
z&=&S\exp(-t \rho \tan(\phi) ),~~~ \phi \in (0,\pi), \phi \ne \frac{\pi}{2},\nonumber\\
w&=&\frac{\epsilon}{\rho\sin(\phi)}\ln S+\frac{1}{\rho}S -V,
~~\epsilon = \pm 1.\nonumber
\end{eqnarray}
Using these expressions as the independent and dependent variables
we reduce the original equation to the ODE
\begin{equation}\label{redh4r0}
- \rho\tan (\phi) w_z
z+\frac{1}{2}\sigma^2(w_{zz}z^2+a)\left[\frac{\alpha( w_z
z-a)}{\alpha( w_z z- a)+a+w_{zz}z^2}\right]^2=0,
 \end{equation}
where $a=(\epsilon\rho\sin(\phi))^{-1}$ and $\phi \in (0,\pi),
(\phi) \ne \frac{\pi}{2}$. Eq. (\ref{redh4r0}) is possible to
reduce to the first order ODE
\begin{equation}\label{firstredh4r0}
  (v+a)\left(z v_z  -(1- \alpha) v + a \right)^2
-\frac{\sigma^2\alpha^2}{2 \rho\tan (\phi)} v^2 (z v_z -v +a)=0
\end{equation}
after substitution $v(z)=z w_z-a$. It is a quadratic algebraic
equation on the value $z v_z$ which roots depends on $v$ only. If
we denote the roots as $f_{\pm}(v)$ we represent the solutions to
Eq. (\ref{firstredh4r0}) in the parametric form
\begin{equation}\label{solfirstredh4r0}
\int{\frac{{\rm d} v}{f_{\pm}(v)}}=\ln{z} +c_1, ~~c_1 \in {\mathbb
R }.
\end{equation}

\subsection{Symmetry reductions in the case $r\ne 0$}

{\begin{theorem} Eq. (\ref{sipar}) admits a three dimensional Lie
algebra $L_3$ with the following operators
\begin{equation}\label{generatorsSiPar}
e_1=\frac{\partial}{\partial t},~~e_2=\left( S - \rho \right)
\frac{\partial}{\partial u},~~ e_3= S \frac{\partial}{\partial S}
+ u  \frac{\partial}{\partial u}.
\end{equation}
The algebra $L_3$ is abelian.
\end{theorem}
Similar to the previous investigation for $r=0$ first we find an
optimal systems of subalgebras for the algebra $L_3$. Because we
are interested just in the one-dimensional subalgebras we provide
the optimal system for these subalgebras only
\begin{eqnarray} \label{optsyssipar}
h_1=<e_3>,\;\;h_2=<e_2 +x e_3>,\:\:h_3=<e_1+x e_2+y e_3 >.
\end{eqnarray}
First we provide for each of these three one-dimensional
subalgebras $h_i, ~i=1,2,3$ and the corresponding subgroup $H_i,
~i=1,2,3$ a set of invariants. Then we use the invariants as the
new independent and independent variables and reduce Eq.
(\ref{sipar}) to some ODEs. In the cases where it is possible we
solve the ODEs.\\

{\bf Case $H_1$}. The algebra $h_1$ is spanned by
\begin{equation}
h_1=<S \frac{\partial}{\partial S} + u  \frac{\partial}{\partial
u}>.
\end{equation}
It describes scaling symmetry of Eq. (\ref{sipar}) and means that
if we multiply the variable $S$ and $u$ with one and the same non
vanishing constant the equation will be unaltered. Respectively
the invariants of this transformations are
\begin{equation}
z=t, ~~ w=\frac{u}{S}.
\end{equation}
If we use these expressions  as the new invariant variables we
obtain a rather simple reduction of the original equation
\begin{equation}
w_z=0,
\end{equation}
with the trivial solution $w=c_1=const.$ It describes all
solutions to Eq. (\ref{sipar}) of the type
\begin{equation}
u(S,t)=Sw(z)=Sw(t)=S c_1.
\end{equation}

{\bf Case $H_2$}. The second subalgebra $h_2$ in the optimal
system of subalgebras  (\ref{optsyssipar}) is given by
\begin{equation}
h_2=< x~ S\frac{\partial}{\partial
S}+\left(S+u(x-\rho)\right)\frac{\partial}{\partial u}
>, ~~x \in {\mathbb R}.
\end{equation}
The invariants of the subgroup $H_2$ are
\begin{eqnarray} \label{invh2sipar1}
z=t,~~
w=-\frac{S^{\frac{\rho}{x}}}{\rho}+S^{\frac{\rho}{x}-1}u,~x\ne 0.
\end{eqnarray}
If we use the invariants (\ref{invh2sipar1}) as the new dependent
and independent variables then the  reduced Eq. (\ref{sipar})
takes the form
\begin{equation}
w_z- \rho \left( \frac{\sigma^2 \alpha^2 (x -\rho)}{2 (\alpha x -
\rho )^2} + \frac{r}{x}\right)w=0.
\end{equation}
It has the solution
\begin{equation}
w(z)=c_1 \exp { \left( \rho \left( \frac{\sigma^2 \alpha^2 (x
-\rho)}{2 (\alpha x - \rho )^2} +\frac{r}{x}\right)\right) t},
~~c_1,x \in {\mathbb R}, x\ne 0.
\end{equation}
The corresponding solution to Eq. (\ref{sipar}) in terms of
variables $S,t,u$ is given by
\begin{equation}
u(S,t)= c_1 S^{1-\frac{\rho}{x}}e^{\rho \gamma t} +\frac{1}{\rho}
S, ~~~~\gamma= \frac{\sigma^2 \alpha^2 (x -\rho)}{2 (\alpha x -
\rho )^2} +\frac{r}{x},
\end{equation}
where $c_1,x \in {\mathbb R}, x\ne 0$ are arbitrary constants.

{\bf Case $H_3$}. The last subalgebra $h_3$ from the optimal
system (\ref{optsyssipar}) is spanned by
\begin{equation}
h_3=<\frac{\partial}{\partial t}+y~S\frac{\partial}{\partial
S}+\left(x~S-x\rho u+y~u\right)\frac{\partial}{\partial u}>,~~ x,y
\in {\mathbb R} .
\end{equation}
We take the two invariants $z,w$
\begin{eqnarray}
z=Se^{-ty},~~w=uS^{\kappa \rho -1}-\frac{1}{\rho}S^{\kappa \rho},~
y\ne 0 ,~~\kappa=\frac{x}{y} .\nonumber
\end{eqnarray}
as  the new independent and dependent variables reduce Eq.
(\ref{sipar}) to the ODE
\begin{eqnarray}
&{}&(r-y)zw_z -{r\kappa \rho} w\label{h3ne0}\\
&+&\frac{\alpha^2\sigma^2}{2}\frac{(z^2w_{zz}+2(1-\kappa\rho)zw_z-\kappa\rho(1-\kappa\rho)w)(zw_z+(1-\kappa
\rho)w)^2}{(z^2w_{zz}+(\alpha +2(1-\kappa \rho)zw_z+(1-\kappa
\rho)(\alpha-\kappa \rho)w)^2}=0. \nonumber
\end{eqnarray}
Eq. (\ref{h3ne0}) possesses a solution of the type
\begin{equation}
w(z)=c_1 z^{q}, ~~c_1 \in \mathbb R,
\end{equation}
where $q$ is a real root of the fifth order algebraic equation
\begin{eqnarray}
((r-y)q -{r\kappa \rho}){(q(q-1)+(\alpha +2(1-\kappa
\rho)q+(1-\kappa \rho)(\alpha-\kappa \rho))^2}  \nonumber\\
+\frac{\alpha^2\sigma^2}{2}
{(q(q-1)+2(1-\kappa\rho)q-\kappa\rho(1-\kappa\rho))(q+(1-\kappa
\rho))^2} =0.\label{h3ne0q}
\end{eqnarray}
Respectively the  solution to Eq. (\ref{sipar}) takes in this case
the form
\begin{equation}
w(z)=c_1 S^{q+1-\kappa \rho}e^{-y q t} +\frac{1}{\rho}S,
~~~c_1,y,\kappa \in \mathbb R, y\ne 0.
\end{equation}

\section{Conclusions}
\label{intr}
 \setcounter{equation}{0}
In the previous sections we studied the Sircar-Papanicolaou model
(\ref{sircpapgen}) in the case of the nonlinear demand function
$U(z)=\beta z, ~ \beta>0 $. The model which includes the linear
demand function was studied in \cite{bib:papanicolaou-sircar-98}
with numerical methods. In this paper we use the methods of Lie
group analysis which gives us a general point of view on the
structure of this equation. We found the symmetry algebra admitted
by the nonlinear PDE (\ref{sipar}) for $r\ne 0$ and by Eq.
(\ref{sipa}) in the case $r=0$. We present in the both cases the
optimal systems of subalgebras. Using the optimal systems of
subalgebras we provide the complete set of non-equivalent
reductions.  In most cases we solve the ODEs or present particular
solutions to them and respectively to Eqs. (\ref{sipar}) and
(\ref{sipa}). The explicit and parametric solutions can be used as
benchmarks for numerical methods.

\end{document}